\documentclass[twocolumn,nofootinbib,superscriptaddress]{revtex4-1}
\usepackage{graphicx,epsfig}
\usepackage{amsmath}
\usepackage{amsfonts}
\usepackage{braket}
\usepackage{fancyhdr}
\newcommand{\be}{\begin{eqnarray}}
\newcommand{\ee}{\end{eqnarray}}

\begin{document}





\title{Remnants in two-dimensional quantum gravity}
\author{Cristiano Germani}
\address{Institut de Ci\`encies del Cosmos, Universitat de Barcelona, Mart\'i i Franqu\`es 1, 08028 Barcelona, Spain}
\author{David Pere\~niguez}
\address{Departament de F\'isica Qu\`antica i Astrof\'isica, Facultat de F\'isica, Universitat de Barcelona, Mart\'i i Franqu\`es 1, 08028 Barcelona, Spain}
\address{DAMTP, Centre for Mathematical Sciences, Wilberforce Road, Cambridge CB3 0WA, UK}

\begin{abstract}
In this work we consider a two-dimensional quantum black hole sourced by the trace anomaly of a conformal field theory. By using holography, we are able to prove that the black hole size is always proportional to the number of states inside the black hole, a result that might be interpreted as a two-dimensional version of the Bekenstein entropy law. Finally, we also show that such a black hole has a minimal size (a remnant). Extrapolating this result for higher dimensions, we show that this would imply that the remnant has a size way larger than the Planck length and is, therefore, always weakly coupled.  
\end{abstract}

\maketitle


\section{Introduction}
Below the Planck energy, quantum gravity fluctuations may be ignored and the spacetime metric is determined by the semiclassical Einstein equations \cite{BirrelDavies}
\begin{equation}\label{eq:1}
R_{\mu\nu}-\frac{1}{2}Rg_{\mu\nu}+\Lambda g_{\mu\nu}=\kappa^{2}_{d}\bra{\psi}T_{\mu\nu}\ket{\psi},
\end{equation}
where $\ket{\psi}$ is the vacuum state of the system and $\Lambda$ is a cosmological constant. The constant $\kappa^{2}_{d}\equiv\left(\ell_p^{(d)}\right)^{d-2}$, where $\ell_p^{(d)}$ is the Planck length in $d$ dimensions \cite{BirrelDavies}. 

The two-dimensional case is very special. There, the Einstein tensor $G_{\mu\nu}=R_{\mu\nu}-\frac{1}{2}Rg_{\mu\nu}$ vanishes identically and $\kappa_{2}$ is dimensionless. Then, \eqref{eq:1} reduces to
\begin{equation}\label{eq:2}
\Lambda g_{\mu\nu}=\kappa_2^2\bra{\psi}T_{\mu\nu}\ket{\psi}\ .
\end{equation} 
A black hole solution \eqref{eq:2} with a conformal field theory (CFT) can be found by inserting a mass $\mu$. In the ``Schwarzschild'' gauge it reads \cite{QuantumBH}
\begin{equation}\label{eq:3}
ds^{2}=-(\lambda^{2}x^{2}+2\mu\lvert x \rvert-1)dt^{2}+\frac{dx^{2}}{\lambda^{2}x^{2}+2\mu\lvert x \rvert-1},
\end{equation}
where $\lambda^{2}$ is related to the trace anomaly of the CFT \cite{BirrelDavies}
\be
\langle T\rangle=-c R\ ,
\ee
via the relation 
\be\label{l1}
\lambda^2=\frac{48\pi\Lambda}{{\cal N}-1}\kappa_2^{-2}\ ,
\ee
where the central charge of the CFT $\lvert c\rvert=\frac{{\cal N}-1}{24\pi}$ \footnote{Note that, as explained in \cite{many}, in the presence of a black hole, the central charge inverts sign from the inside to the outside. This might also be interpreted that the interior of a black hole is a condensate of the CFT \cite{dvali}. See also \cite{cunillera, giugno} for recent developments.}, ${\cal N}$ is the number of degrees of freedom of the CFT. 

We will here only consider the case $\Lambda>0$. Note that in the absence of the mass $\mu$, the metric  \eqref{eq:2} would be the one of an anti de Sitter space in two-dimensions.  

The fact that in two dimensions $k_2$ is dimensionless, precludes us to define a scale at which the semiclassical approximation \eqref{eq:1} breaks down. From a path integral perspective, this means that there is no any energy scale below which the classical gravitational saddle point dominates over others, and so this has to be checked case by case, dependently on the observable chosen. Quite generically though, since in two dimensions the Einstein-Hilbert action is a total derivative, any correlation functions from the path integral can be calculated by ``only'' considering topological (gravitational) diversities, real and complex, with the same boundary conditions. 

A step towards this full integration has been done in \cite{many}. There, it was noticed that the quantum system described above, with semiclassical real metric \eqref{eq:3}, is a combination of two Liouville field theories, a timelike (the black hole interior) and a spacelike (the black hole exterior) matched at the black hole horizon. In the same paper, it was then shown that observables living inside the black hole and described by heavy operators (i.e. semiclassical operators) are only correctly calculated by considering two distinct saddle points, one real, corresponding to the metric \eqref{eq:3}, and one complex. Unfortunately, given the high complexity of Liouville field theories any step forwards would be an incredibly hard task. For this reason we will take an alternative route: holography.

In the Maldacena AdS/CFT conjecture \cite{maldacena} gravity in $d+1$ dimensions is dual to a CFT living at the (Minkowskian) boundary of AdS. Any other slice of AdS space, away from its boundary, contains a local gravitational field. Therefore, cutting the AdS space at any slice away from the AdS boundary, immediately introduces gravity in the dual CFT. Of course, this non trivial cut comes with the price of introducing a local energy (a tension) of the cut. Precisely this tension parameterises the local gravitational strength and therefore the UV cut-off of the theory. In this respect, the AdS boundary is a zero tension brane so that the effective gravitational strength vanishes and the CFT has no cut-off. 

The geometrical construction of cutting and pasting an AdS space is the so-called Randall-Sundrum II braneworld (RS) \cite{rs} and the new ``AdS'' boundary is called {\it the brane}. 

One of the first interesting result of this variant of the AdS/CFT has been found by a re-interpretation of the results of \cite{collapse} by \cite{emparan}. In \cite{collapse}, it has been proven that a gravitational collapse of a ball of dust, stuck on a RS brane, would not end into a Schwarzschild black hole, as it would happen in classical General Relativity. Rather, a ball of dust would generate a time dependent solution sourced by a stress tensor matching the trace anomaly of the would-be black hole. Motivated by this result, the authors of \cite{emparan} have then conjectured that
\\\\
{\centering{\textit{\footnotesize "The black hole solutions localised in the brane in the AdS$_{d+1}$ braneworld which are found by solving the classical bulk equations in AdS$_{d+1}$ with the brane boundary conditions, correspond to quantum-corrected black holes in $d$ dimensions, rather than classical ones."}}}
\\\\
In \cite{anti}, this conjecture has been tested further. There, strong evidences are given towards the interpretation of the result of \cite{collapse} in terms of a Hawking flux of the would be Schwarzschild black hole. 

In \cite{2DimQBH}, it has been shown that the quantum metric \eqref{eq:3} is found by brane slicing a BTZ black hole. This brane is formed by two intersecting strings which can be the (dual) reason why quantum correlation functions of heavy operators on the black hole \eqref{eq:3}, are given by two saddle points of the gravitational path integral, as discussed in \cite{many}. 

In this paper we take again seriously the conjecture of \cite{emparan} and show that, quite un-expectedly, the two dimensional black hole studied in \cite{2DimQBH} has a minimal size, a remnant.
\section{1-brane black hole in BTZ bulk}
In this section we review and extend the analysis of \cite{2DimQBH}. 

A spherically symmetric asymptotically AdS black hole solution in 2+1 dimensions, solution of Einstein gravity, is the BTZ black hole
\begin{equation}\label{eq:4}
ds^{2}=-F(r)dt^{2}+\frac{dr^{2}}{F(r)}+r^{2}d\theta^{2}\ ,
\end{equation}
where $F(r)=\frac{r^{2}}{L^{2}}-m$, $L$ is the AdS$_{3}$ length, $\mathcal{M}\equiv m/L$ is the ``mass'' of the black hole. Finally, conversely to AdS$_3$, the coordinate $\theta$ is identified as $\theta \sim \theta+2\pi$ so that $r=0$ cannot be analytically extended. For this reason  the ``singularity'' at $r=0$ of \eqref{eq:4}, hidden by the horizon at $r_h=\sqrt{m}L$, is not a singularity in the curvature, but rather a point of geodesic incompleteness (i.e. a causal singularity). From the perspective of AdS$_3$, that is extendible to negative values of $r$, the identification of $\theta$ would generate closed time-like curves and therefore must be discarded \cite{BTZG}. In short, BTZ is only defined in the coordinate range $r> 0$.

We are now interested in constructing a braneworld with a BTZ bulk. To do so, we will cut the BTZ geometry by a 1-brane which will act as the boundary of our new spacetime. More geometrically, we will first slice the BTZ space into two parts delimited by a 1-brane. We will then select one them to be copied and pasted on the other side of the brane such that the final space is ${\cal Z}_2$ symmetric with respect to the brane.

Calling $\Sigma$ the brane, we have that our braneworld is described by the action
\begin{multline}\label{eq:7}
\mathcal{S}[g_{\alpha\beta},h_{\alpha\beta}]=\frac{1}{2\kappa_{3}^{2}}\int d^{3}x\sqrt{-g}(R+\frac{2}{L^{2}})+\\
+\frac{1}{\kappa_{3}^{2}}\int_{\Sigma}d^{2}y\sqrt{-h}(K^{+}+K^{-})-\frac{2\sigma}{\kappa_{3}^{2}}\int_{\Sigma}d^{2}y\sqrt{-h}\ ,
\end{multline} 
where $1/\kappa_{3}^{2}$ is the 3-dimensional Planck mass, $K^\pm$ are the extrinsic curvatures with respect to the brane calculated on the right and on the left of the brane itself. Finally, $h_{\alpha\beta}$ is the induced metric on the brane and $\sigma$ is the brane's tension. 

The variation w.r.t. $g^{\alpha\beta}$ is nothing else than the Einstein equations which are solved by \eqref{eq:4} while the variation w.r.t $h^{\alpha\beta}$ gives
\begin{equation}\label{eq:12}
K_{\alpha\beta}=\sigma h_{\alpha\beta}.
\end{equation}
where, given the ${\cal Z}_2$ symmetry, $K\equiv K^+$.

As in \cite{2DimQBH} we consider the following profile for the 1-brane 
\begin{equation}\label{eq:13}
\Sigma : \,\, 0=\Theta(r,\theta):=\theta-\Psi(r),\,\,\,\,\,\Psi\sim\Psi+2\pi\ .
\end{equation}
In order to construct $K_{\alpha\beta}=\frac{1}{2}\pounds_{n} g_{\alpha\beta}$ we need the normal vector $n^\alpha$ to the 1-brane. This is 
\be
n_{\alpha}=\frac{\pm\partial_{\mu}\Theta}{\sqrt{\lvert\partial_{\mu}\Theta\partial^{\mu}\Theta\rvert}}=\pm A(0,-\Psi'(r),1)\ ,
\ee 
where $\Psi '(r):=\frac{d}{dr}\Psi(r)$ and $A=\frac{r}{\sqrt{F(r)(r\Psi'(r))^{2}+1}}$. 

\begin{figure}[h!]
\centering
\includegraphics[scale=0.4, angle=-90]{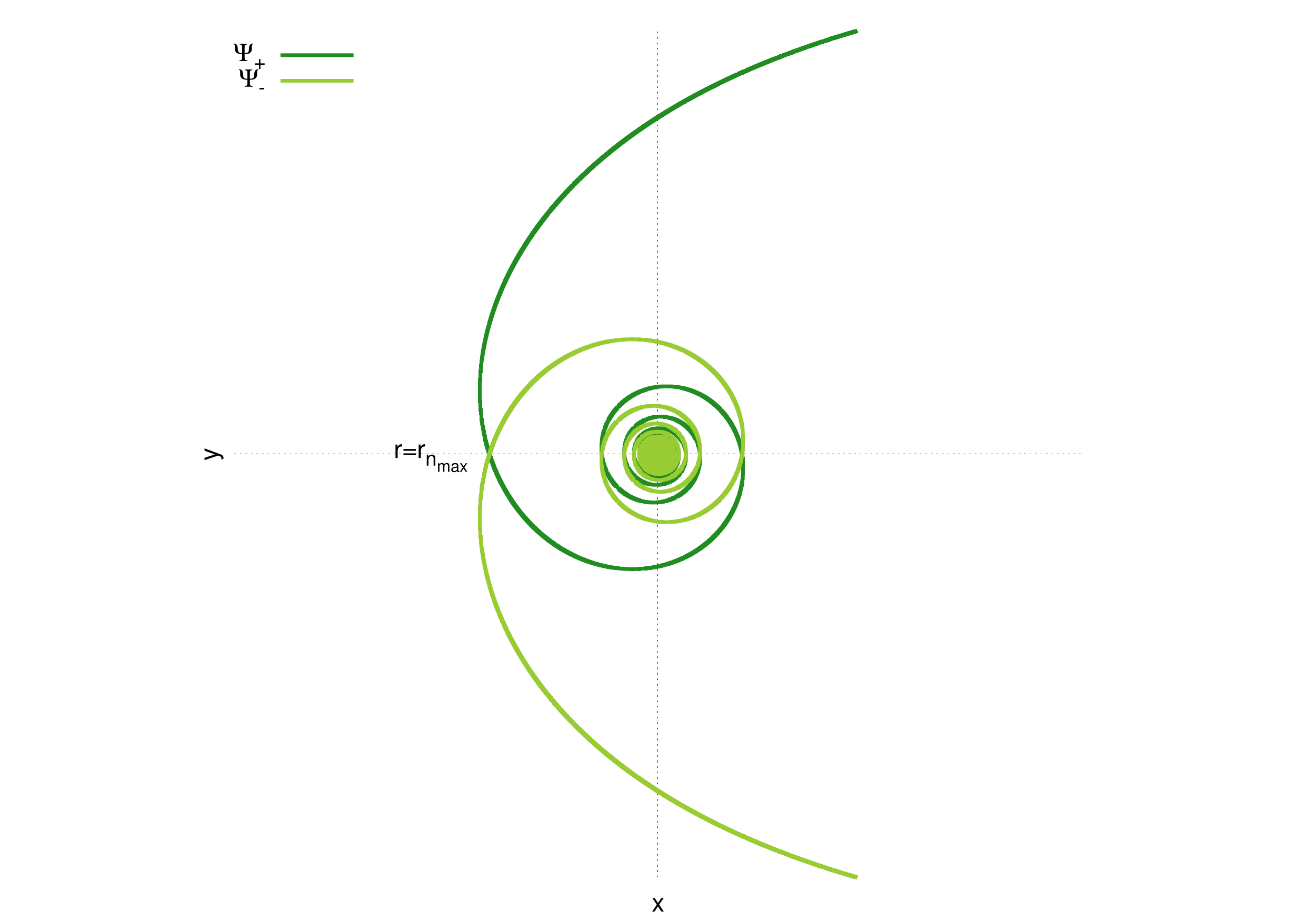}
\caption{Schematic representation of the strings wrapping around the centre of the BTZ black hole.}
\label{fig0}
\end{figure}

With this, equations \eqref{eq:12} are simultaneously solved by the following two profiles 
\begin{equation}\label{eq:16}
\Psi_{\pm}(r)=\pm\frac{\log\biggl(\frac{2\sigma^{2}L^{4}m+2\sigma L^{2}\sqrt{m}\sqrt{r^{2}(1-\sigma^{2}L^{2})+\sigma^{2}L^{4}m}}{Lr}\biggr)}{\sqrt{m}}.
\end{equation}
These two solutions correspond to two strings symmetric w.r.t. the $x$-axis wrapping around the centre of the space an infinite number of times (see fig. \eqref{fig0}). 

From eq. \eqref{eq:16} we see that one has to impose $\sigma^{2}L^{2}< 1$ for a real solution of an infinitely large brane.
The induced metric on either $\Psi_{+}$ or $\Psi_{-}$ is 
\begin{equation}\label{eq:18}
ds^{2}=-\biggl(\frac{r^{2}}{L^{2}}-m\biggr)dt^{2}+\frac{\phi(r)}{\frac{r^{2}}{L^{2}}-m}dr^{2}\ ,
\end{equation}
being $\phi(r)=\frac{2\alpha r^{2}}{L(\beta r^{2}+\alpha^{2})}$, $\alpha=2\sigma^{2}L^{3}m$ and $\beta=4\sigma^{2}L^{2}m(1-\sigma^{2}L^{2})$. 

Changing into Schwarzschild-like coordinates we have 
\be ds^{2}=-\tilde{F}(\rho)dt^{2}+\frac{d\rho^{2}}{\tilde{F}(\rho)},
\ee
with 
\be
\tilde{F}(\rho)=\frac{1-\sigma^{2}L^{2}}{L^{2}}\rho^{2}-\frac{m}{(1-\sigma^{2}L^{2})}\ ,
\ee
and 
\begin{equation}
\rho(r)=\frac{\sqrt{(1-\sigma^{2}L^{2})r^{2}+\sigma^{2}L^{4}m}}{1-\sigma^{2}L^{2}}\ .
\end{equation}

\begin{figure}[h!]
\centering
\includegraphics[scale=0.4]{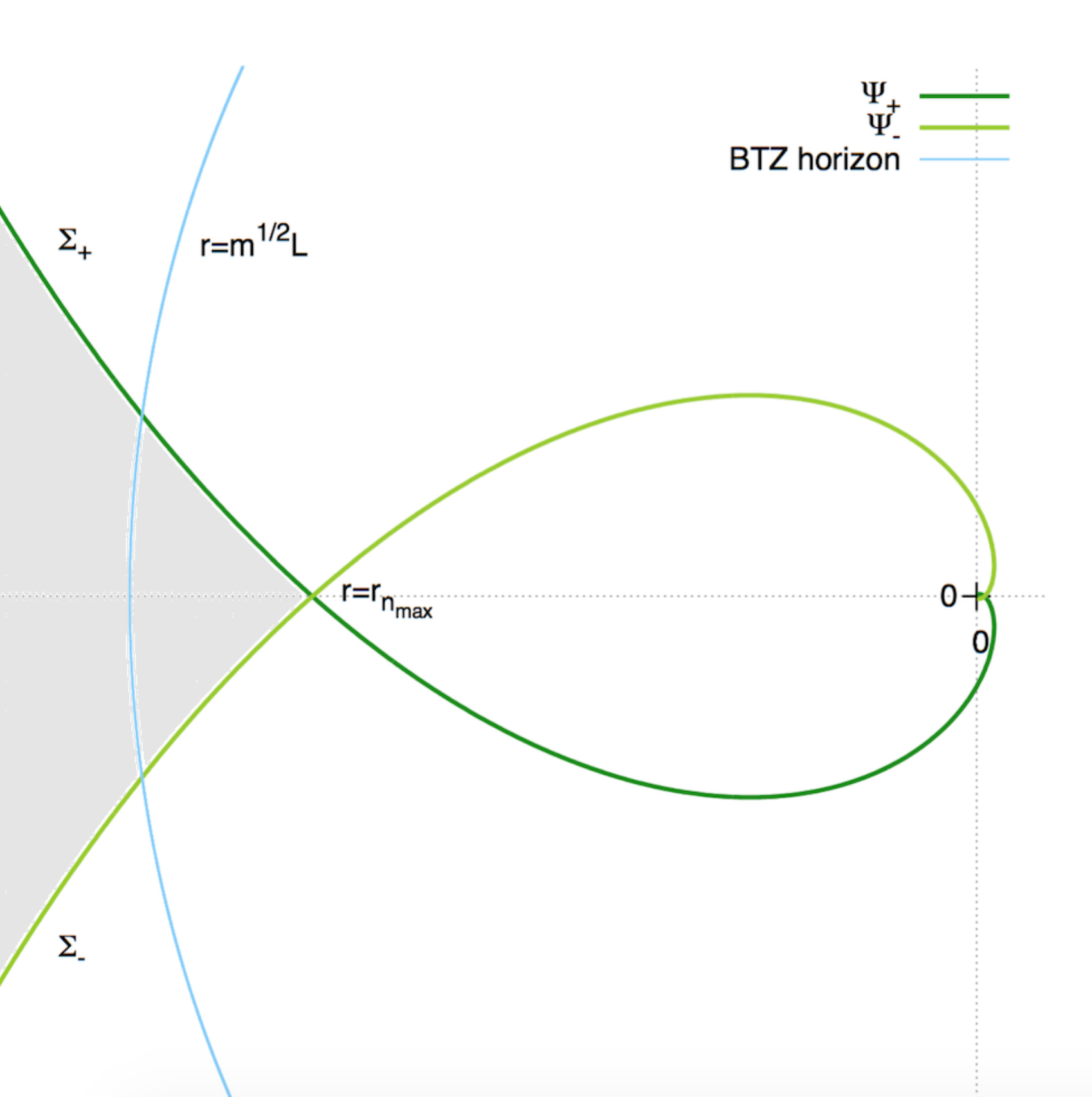}
\caption{$\Psi_{\pm}$ solutions. The shaded region is the braneworld.}
\label{fig1}
\end{figure}

To construct a braneworld we however need to cut the space into two separate parts. Since each one of the strings wrap around an infinite amount of times we cannot use neither of those to be the boundary of our spacetime. Then, a way to construct a 2-dimensional brane black hole is to consider the space delimited by the first intersection of the two strings, as depicted in fig.\eqref{fig1}. 
The geometry of two intersecting strings is however the same of a pinched string. The pinching can be re-interpreted as an insertion of a point mass in the position of the pinching. Thus, this construction will be a solution of \eqref{eq:7} with an additional boundary mass term.

The radius at which $\Psi_{+}$ and $\Psi_{-}$ intersect are given by
\begin{equation}\label{eq:24}
r_{n}=L\frac{4\sigma^{2}L^{2}e^{n\pi\sqrt{m}}m}{e^{2n\pi\sqrt{m}}-4\sigma^{2}L^{2}(1-\sigma^{2}L^{2})m}\ ,
\end{equation}
with $n\in\mathbb{Z}$ and $\Psi_{\pm}$ approaching infinity with an asymptotic angle
\begin{equation}\label{eq:25}
\theta_{\infty}(\sigma^{2})=\pm \frac{1}{2\sqrt{m}}\log\bigl(4\sigma^{2}L^{2}(1-\sigma^{2}L^{2})m\bigr).
\end{equation}
Indeed, when $\theta_{\infty}(\sigma^{2})/\pi\in\mathbb{Z}$ the branches become parallel for large $r$, the denominator in \eqref{eq:24} vanishes for $n=\theta_{\infty}(\sigma^{2})/\pi$ and therefore the last intersection point lies at infinity. It is easy to check that $\frac{dr_{n}}{dn}<0\,\,\,(\forall\sigma,m,n)$, so when $\theta_{\infty}(\sigma^{2})/\pi\notin\mathbb{Z}$ there exists a last intersecting point given by $r_{n_{max}}$, being
\begin{equation}\label{eq:26}
n_{max}(\sigma^{2})=\biggl[\frac{1}{\pi}\theta_{\infty}(\sigma^{2})\biggr]_{\rightarrow}\ ,
\end{equation}
where the operator $[\cdot]_{\rightarrow}$ takes the first integer coming after the $\mathbb{R}$-number it contains. For $r>r_{n_{max}}$ the branches do not intersect and approach infinity with asymptotic angle given by \eqref{eq:25}. 

Finally, the 1-brane of our braneworld ($\Sigma$) can be constructed by cutting out the parts of $\Psi_{+}$ and $\Psi_{-}$ at which $r<r_{n_{max}}$, and gluing the remaining branches, that we will refer to as $\Sigma_{+}$ and $\Sigma_{-}$ respectively, in the last intersection point $r=r_{n_{max}}$ (see fig.\eqref{fig1}). 

Let us first study the branch $\Sigma_{+}$. Introducing the coordinate $x:=\rho-\rho(r_{n_{max}})$, $0<x<\infty$, the metric on $\Sigma_{+}$ reads
\begin{equation}\label{eq:27}
ds^{2}_{\Sigma_{+}}=-(\lambda^{2}x^{2}+2Mx-N)dt^{2}+\frac{dx^{2}}{\lambda^{2}x^{2}+2Mx-N}\ ,
\end{equation} 
where 
\be
\lambda^{2}&=&\frac{1-\sigma^{2}L^{2}}{L^{2}}\ ,\cr 
M&=&\frac{1}{L}\sqrt{\frac{r_{n_{max}}^{2}}{L^{2}}(1-\sigma^{2}L^{2})+\sigma^{2}L^{2}m}\ ,\cr
N&=&m-\frac{r_{n_{max}}^{2}}{L^{2}}\ .
\ee 
Doing a similar procedure on $\Sigma_{-}$, but defining $x:=-(\rho-\rho(r_{n_{max}}))$, $-\infty<x<0$ instead, allows us to write down the induced metric on $\Sigma$ in the compact form
\begin{equation}\label{eq:28}
ds^{2}_{\Sigma}=-f(x)dt^{2}+\frac{dx^{2}}{f(x)}\ ,
\end{equation}
where $f(x):=\lambda^{2}x^{2}+2M\lvert x\rvert-N$. 

The horizon of \eqref{eq:28} lies at
\begin{equation}\label{eq:29}
\lvert x_{h}\rvert=\frac{-M+\sqrt{M^{2}+\lambda^{2}N}}{\lambda^{2}},
\end{equation}
and it will only exist if $N>0$, i.e. if the last intersecting point is inside the BTZ horizon. In other words, we need $r_{n_{\max}}<\sqrt{m}L$. Finally, as in this construction $N>0$, we can rescale the coordinates as $\tilde{x}=\frac{x}{\sqrt{N}}$ and $\tilde{t}=\sqrt{N}t$, and the metric becomes
\begin{equation}\label{eq:30}
ds^{2}_{\Sigma}=-\tilde{f}(\tilde{x})d\tilde{t}^{2}+\frac{d\tilde{x}^{2}}{\tilde{f}(\tilde{x})}\ ,
\end{equation}
where $\tilde{f}(\tilde{x})=\lambda^{2}\tilde{x}^{2}+2\mu\lvert \tilde{x}\rvert-1$, $\mu=\frac{M}{\sqrt{N}}\geq 0$ (from now on we remove the tildes). This is the same quantum metric \eqref{eq:3} obtained from the two-dimensional gravity coupled to a CFT. However, here a comment is necessary. The braneworld relation between the three-dimensional and the two-dimensional Planck scale is $\kappa_3^2\sim \frac{k_2^2}{\sigma}$ \cite{sasaki}. As we are using a classical bulk $\kappa_3^2<\infty$. Then, the limit $\sigma\rightarrow 0$ would also imply $k_2^2\rightarrow 0$, i.e. absence of gravity in the two-dimensional theory. Therefore, in order to interpret our braneworld metric \eqref{eq:30} as the two-dimensional metric from the semiclassical gravity equations \eqref{eq:2}, we would need to avoid the $\sigma\rightarrow 0$ limit. 

So far, we have reviewed \cite{2DimQBH}, in the following we will analyse further the brane solution \eqref{eq:30} and show that, in the limit of large bulk black hole mass, the brane black hole has, at leading order in $m$, a minimal size. Thus, if \eqref{eq:30} holographically represents the quantum two-dimensional black hole as in the \cite{emparan} conjecture, the two-dimensional quantum black hole has a remnant. 

\section{Analysis of the Black Hole Solution on $\Sigma$}
\begin{figure}[h!]
\centering
\includegraphics[scale=0.3]{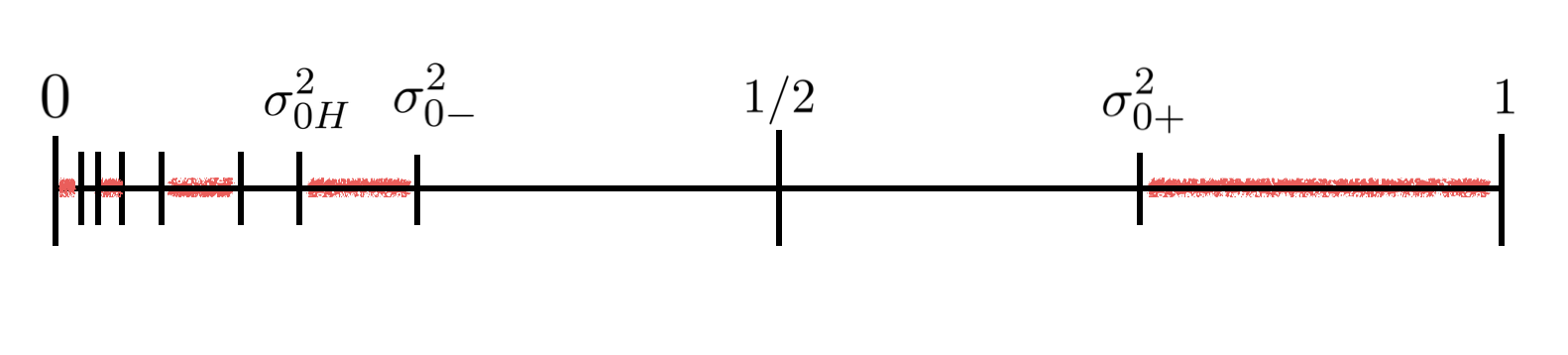}
\caption{In red the interval of the tensions in which no brane black hole can be constructed. $L=1$ units are chosen here.}
\label{fig3}
\end{figure}

The use of a classical BTZ solution in the bulk makes only sense for a mass much larger than the inverse AdS length. In addition, for a classical definition of an AdS length, $L$ must be much longer than the Planck length. This comes in the following hierarchies of scales
\be
m\gg 1\ ,\ L\gg k_3^{-2}\ .
\ee 
A necessary condition to have the black hole on $\Sigma$ is that the last intersecting point between the branes must lie inside the BTZ horizon, that is
\begin{equation}\label{eq:31}
r_{n_{\text{max}}}<\sqrt{m}L.
\end{equation}
The condition \eqref{eq:31} for $m>1$ is {\it violated} in the intervals
\begin{equation}\label{eq:32}
 \bigl(\cup_{n=-\infty}^{0}[\sigma_{nH}^{2},\sigma_{n-}^{2}]\bigr)\cup[\sigma^{2}_{0+},L^{-2}] 
\end{equation}
where
\begin{equation}\label{eq:33}
\sigma^{2}_{n\pm}=\frac{1}{2L^{2}}\biggl(1\pm\sqrt{1-\frac{\exp{(2n\pi\sqrt{m})}}{m}}\biggr)\ ,
\end{equation}
and 
\begin{multline}\label{eq:34}
\sigma^{2}_{nH}=\frac{1}{2L^{2}}\Biggl(1+\frac{\exp{(n\pi\sqrt{m})}}{\sqrt{m}}-\\
-\sqrt{\biggl(1+\frac{\exp{(n\pi\sqrt{m})}}{\sqrt{m}}\biggr)^{2}-\frac{\exp{(2n\pi\sqrt{m})}}{m}}\Biggr).
\end{multline}
Where \eqref{eq:33} and \eqref{eq:34} are respectively solutions of
\begin{equation}\label{eq:35}
i)\,\,n=\frac{1}{\pi}\theta_{\infty}(\sigma^{2}_{n\pm})\,\,\,\forall\ n\leq0\ ,\ ii)\,\, r_{n}(\sigma^{2}_{nH})=\sqrt{m}L\ .
\end{equation}
Therefore, $\sigma^{2}_{n\pm}$ are the tensions at which $r_{n}$ lies at infinity, and $\sigma^{2}_{nH}$ the tension at which $r_n$ lies at the BTZ horizon. The result \eqref{eq:32} is schematically represented in fig. \eqref{fig3}, and has a clear geometrical interpretation:
\begin{itemize}
\item Whenever $\sigma^{2}\in(\sigma^{2}_{0-},\sigma^{2}_{0+})$, the last intersecting point is $r_{1}$, and $r_{1}<\sqrt{m}L\,\,\,\forall\sigma^{2}\in(0,1)$. That is, $r_{1}$ is always inside the horizon. Therefore, the black hole is allowed in $(\sigma^{2}_{0-},\sigma^{2}_{0+})$.
\item For $\sigma^{2}\in[\sigma_{0+}^{2},1)$, the branes bend in such a way that the intersecting point $r_{0}$ appears from infinity and becomes the last intersecting point. In that interval of tensions, $r_{0}>\sqrt{m}L$ and, hence, there can not be a black hole on the brane.
\item For $\sigma^{2}\in(0,\sigma^{2}_{0-}]$, the branes bend again and a new intersecting point appears from infinity. In that situation the black hole on $\Sigma$ can only be constructed whenever that point enters the horizon. Subsequent lowering of the tension would again produce new bending and therefore new intersecting points coming from infinity. This process leads to the spectrum of forbidden intervals shown in \eqref{eq:32}. 
\end{itemize}
As we have already mentioned, our brane solution only makes sense in the limit $m\gg1$. In BTZ, this limit also means high temperature as $T_{BTZ}\sim\sqrt{m}$ \cite{BTZ}. We will then analyse the brane geometry found before in the large $T_{BTZ}$ limit by expanding in the small parameter $\epsilon\equiv 1/\sqrt{m}$. In particular, we are interested in studying the size of the black hole horizon. Defining the dimensionless parameter $\tilde{\mu}\equiv\mu/\lambda$, the horizon of our 1-brane black hole is given by
\begin{equation}\label{eq:37}
\lvert x_{h}\rvert=\frac{-\tilde{\mu}+\sqrt{\tilde{\mu}^{2}+1}}{\lambda}\ .
\end{equation} 
Furthermore, expansion in $\epsilon$ of \eqref{eq:32} gives
\begin{equation*}
\frac{1}{L^2} \bigl(\cup_{n=-\infty}^{0}[\mathcal{O}_{n}(\epsilon^{2}),\mathcal{O}_{n}(\epsilon^{2})]\bigr)\cup[1-\frac{\epsilon^{2}}{4}+\mathcal{O}(\epsilon^{3}),1] \ .
\end{equation*}
Therefore, in the large $T_{BTZ}$ limit the central region of fig.\eqref{fig3} is the only one relevant as the region on the left of $\sigma_{0^-}^2$ would not be dual to any quantum gravity (there gravity decouples from the CFT, as explained before). Hence, to study the horizon we restrict to $L^2\sigma^2\in(\frac{\epsilon^2}{4},1-\frac{\epsilon^2}{4})$. We remind the reader that in this interval $n_{\rm max}=+1$. In such region, $\lvert x_{h}\rvert$ is a monotonically decreasing function of $\sigma$ and hence, it exhibits upper and lower bounds. Precisely the opposite happens for $\mu$ as a function of $\sigma$.

In the large $T_{BTZ}$ limit, the bounds on $\lvert x_{h}\rvert$ are given by
\be
\lvert x_{h}^{\rm max}(\frac{\epsilon^2}{4L^2})\rvert_{n_{\rm max}=1}\simeq \frac{1}{\lambda(\frac{\epsilon^2}{4L^2})}\ ,
\ee 
and 
\be
\rvert x_{h}^{\rm min}(L^{-2})\rvert_{n_{\rm max}=1}\simeq \frac{L}{2}\ .\ee
The maximal size give us no surprises as it is exactly equal to the 2-dimensional AdS length. In fact, note that in two dimensions the larger is the mass the smaller the black hole is, conversely to the higher dimensional cases. Therefore, in the limit $\mu\rightarrow 0$ the only horizon surviving is the AdS horizon. Of course strictly speaking a black hole is only such for $\mu\neq 0$ (otherwise it is simply AdS). It is however quite a surprise to see the existence of a minimum size and we will discuss this in the next section.

\section{Conclusions: Constraints on the Quantum Black Hole and Remnant} 

We can now use the AdS/CFT dictionary \cite{Malda}, valid for ${\cal N}\gg 1$, to relate the number of fields in the CFT to the three-dimensional quantities
\begin{equation}\label{eq:40}
{\cal N}=\frac{12\pi L}{\kappa_{3}^{2}}\ ,
\end{equation}
to find
\begin{equation}\label{eq:41}
\lvert x^{\text{min}}_h\rvert={\cal N}\frac{\ell^{(3)}_{p}}{24\pi}\ .
\end{equation}
We can interpret the previous minimal length by defining a universal minimal cell of size $\ell_{min}\equiv \frac{\ell^{(3)}_{p}}{24\pi}$. Then a black hole with $\cal N$ degrees of freedom cannot occupy less space than $\cal N$ times $\ell_{min}$. Note that $\ell_{min}$ does not depend on the BTZ mass and/or the AdS length and therefore it will not change by changing $m$ and/or $L$. If the same result applied to larger dimensions, we would obtain a black hole remnant that would have a size way larger than the quantum gravity scale and thus be weakly coupled (note that our results are valid also for ${\cal N}\sim 24\pi$ where $\lvert x^{\text{min}}_h\rvert\sim \ell^{(3)}_p$). To prove this statement one can use the fact that  $k_{d-1}^2\sim k_d^2\sigma$ and that, to be in the classical braneworld regime, $\sigma\ll k_d^{2/(2-d)}$. Therefore, $\ell_p^{(d)}\gg \ell_p^{(d-1)}$ which proves that the remnant is in the weak gravity regime.

It is interesting to rewrite the minimal radius \eqref{eq:41} in terms of only the two-dimensional constants. One finds that  
\be\label{eq:40n}
\lvert x^{\text{min}}_h\rvert\sim \frac{\cal N}{24\pi\mu_{\text{max}}}\ ,
\ee
where $\mu_{\rm max}=\mu(\sigma=L^{-1})\simeq L^{-1}$.

The entropy of a two-dimensional conformal field theory is given by the Cardy formula \cite{cardy}
\be
S_C=2\pi\sqrt{\frac{c}{6}\left(E R-\frac{c}{24}\right)}
\ee
where, for the remnant,  $E=\mu_{\rm max}$ is the energy \cite{2DimQBH} and $R=\lvert x^{\rm min}_h\rvert$ the size of the system.

From equation \eqref{eq:40n} and by using that $\lvert c\rvert\sim {\cal N}$, we then immediately find that $S_C\sim {\cal N}$ and so 
\be
S_C\sim \frac{\lvert x^{\rm min}_h\rvert}{\ell_{min}}\ .
\ee 
This resembles the Bekenstein entropy law. Namely that a larger black hole has larger entropy and therefore contains larger number of states. 

The fact that the black hole size is proportional to the entropy must also be true for any black hole size and not only for the minimal one. As we have already commented, the maximal radius of our two-dimensional black hole is the two-dimensional AdS length and corresponds to the minimal tension $\epsilon/2$. In this case, one finds 
\be
\lvert x^{\text{max}}_h\rvert=2\lvert x^{\text{min}}_h\rvert\ ,
\ee
and thus, $\lvert x_h\rvert= {\cal O}(1)\lvert x^{\text{min}}_h\rvert$. Therefore,
\be
\lvert x_h\rvert\sim \ell_{min} S_C\ ,
\ee
which is what we were expecting.

\begin{acknowledgments}
The authors would like to thank  Biel Cardona, Francesc Cunillera and Roberto Emparan for useful discussions. CG is supported by the Ramon y Cajal program and partially supported by the Unidad de Excelencia Mar\'ia de Maeztu Grant No. MDM-2014-0369 and by the national FPA2013-46570-C2-2-P and FPA2016-76005-C2-2-P grants. DP is supported by ``La Caixa'' fellowship grant for post-graduate studies and the Spanish MECD ``beca de colaboraci\'on con departamentos''.
\end{acknowledgments}


\begin{thebibliography}{99}
\bibitem{BirrelDavies} N.D. Birrell and P.C.W. Davies, \textit{Quantum fields in curved space}, Cambridge University Press, New York (1982).
\bibitem{QuantumBH} R. B. Mann, A. Shiekh and L. Tarasov, Nucl. Phys. B \textbf{341}, 134 (1990).
\bibitem{many} C.~Germani,
  Phys.\ Lett.\ B {\bf 733} (2014) 93
  doi:10.1016/j.physletb.2014.04.030
  [arXiv:1307.6238 [hep-th]].
\bibitem{dvali} G.~Dvali and C.~Gomez,
  Fortsch.\ Phys.\  {\bf 61} (2013) 742
  doi:10.1002/prop.201300001
  [arXiv:1112.3359 [hep-th]].
\bibitem{cunillera} F.~Cunillera and C.~Germani,
  arXiv:1711.01282 [gr-qc].
\bibitem{giugno}A.~Giugno, A.~Giusti and A.~Helou,
  arXiv:1711.06209 [gr-qc].
\bibitem{maldacena}  J.~M.~Maldacena,
  Int.\ J.\ Theor.\ Phys.\  {\bf 38} (1999) 1113
   [Adv.\ Theor.\ Math.\ Phys.\  {\bf 2} (1998) 231]
  doi:10.1023/A:1026654312961, 10.4310/ATMP.1998.v2.n2.a1
  [hep-th/9711200].
  \bibitem{rs} L.~Randall and R.~Sundrum,
  Phys.\ Rev.\ Lett.\  {\bf 83} (1999) 4690
  doi:10.1103/PhysRevLett.83.4690
  [hep-th/9906064].
\bibitem{collapse} M.~Bruni, C.~Germani and R.~Maartens,
  Phys.\ Rev.\ Lett.\  {\bf 87} (2001) 231302
  doi:10.1103/PhysRevLett.87.231302
  [gr-qc/0108013].
\bibitem{emparan} R.~Emparan, A.~Fabbri and N.~Kaloper,
  JHEP {\bf 0208} (2002) 043
  doi:10.1088/1126-6708/2002/08/043
  [hep-th/0206155].
\bibitem{anti} R.~Casadio and C.~Germani,
  Prog.\ Theor.\ Phys.\  {\bf 114} (2005) 23
  doi:10.1143/PTP.114.23
  [hep-th/0407191].
\bibitem{2DimQBH}  C.~Germani and G.~P.~Procopio,
  Phys.\ Rev.\ D {\bf 74} (2006) 044012
  doi:10.1103/PhysRevD.74.044012
  [hep-th/0605068].
\bibitem{BTZG} M. Ba\~nados, M. Henneaux, C. Teitelboim and J. Zanelli,  Phys. Rev. D \textbf{48}, 1506 (1993).
\bibitem{BTZ} M. Ba\~nados, C. Teitelboim and J. Zanelli, Phys. Rev. Lett. \textbf{69}, 1849 (1992).
\bibitem{sasaki} T.~Shiromizu, K.~i.~Maeda and M.~Sasaki,
  Phys.\ Rev.\ D {\bf 62} (2000) 024012
  doi:10.1103/PhysRevD.62.024012
  [gr-qc/9910076].
\bibitem{Malda} O.~Aharony, S.~S.~Gubser, J.~M.~Maldacena, H.~Ooguri and Y.~Oz,
  Phys.\ Rept.\  {\bf 323} (2000) 183
  doi:10.1016/S0370-1573(99)00083-6
  [hep-th/9905111].
\bibitem{cardy}
J.~L.~Cardy,
  Nucl.\ Phys.\ B {\bf 270} (1986) 186.
  doi:10.1016/0550-3213(86)90552-3
\end{thebibliography}
\end{document}